\documentclass[journal,letterpaper]{IEEEtran_nssmic}

\usepackage{graphicx}  

\begin{document}

\title{Tests of a proximity focusing RICH \\
with aerogel as radiator}

\author{ I.~Adachi, I.~Bizjak, A.~Gori\v{s}ek, T.~Iijima,
 M.~Iwamoto, S.~Korpar, P.~Kri\v zan, R.~Pestotnik,  M.~Stari\v{c},  
A.~Stanovnik,  T.~Sumiyoshi, K.~Suzuki and T.~Tabata 
\thanks{Manuscript received December 1, 2002.}
\thanks{I.~Adachi and K.~Suzuki are with the  High Energy Accelerator 
Research Organization (KEK), Japan. M.~Iwamoto and   T.~Tabata are with the
University of Chiba, Japan.  T.~Sumiyoshi is with the Tokyo Metropolitan 
University, Tokyo, Japan, and T.~Iijima is with the Physics Department, Nagoya 
University, Nagoya, Japan.  P.~Kri\v zan is with the  Faculty of Mathematics 
and Physics, University of Ljubljana, and with the Jo\v{z}ef Stefan Institute, 
Ljubljana, Slovenia. I.~Bizjak, A.~Gori\v{s}ek, R.~Pestotnik and M.~Stari\v{c} 
are with the  Jo\v{z}ef Stefan Institute. S.~Korpar is
with the Faculty of Chemistry and Chemical Engineering, University of Maribor,
and with the  Jo\v{z}ef Stefan Institute. A.~Stanovnik is with 
the Faculty of Electrical Engineering, University of Ljubljana, and 
with the  Jo\v{z}ef Stefan Institute.}}

\maketitle

\begin{abstract}
Using aerogel as radiator and multianode PMTs for photon detection,
a proximity focusing Cherenkov ring imaging detector has been constructed
and tested in the KEK $\pi$2 beam.
The aim is to experimentally study the basic parameters such as resolution
of the single photon Cherenkov angle and number of detected 
photons per ring. The 
resolution obtained is well approximated by estimates of contributions from
 pixel size and emission point uncertainty. 
The number of detected photons per Cherenkov ring is in good agreement with
estimates based on aerogel  and detector characteristics.  The values obtained 
turn out to be  rather low, mainly due to
Rayleigh scattering and to the relatively large dead space between the 
photocathodes. A light collection system or a higher fraction
of the  photomultiplier active area, together with better quality aerogels
are expected to improve the situation. The reduction of Cherenkov yield,
for charged particle impact in the vicinity of the aerogel tile side wall,
has also been measured.

\end{abstract}


\section{Introduction}
\label{In}

Aerogels are materials with density and refractive index in the region between
gases and liquids or solids. Already some time ago, Cantin et al. \cite{1} proposed
that Cherenkov radiation from silica aerogels could be used for detection
of particles. Besides particle  detectors  like for example TASSO 
\cite{2}, such threshold counters found applications also in other 
fields \cite{3}. With improved manufacturing techniques, aerogels of higher
transparency i.e. less Rayleigh scattering became available, permitting
their consideration as radiators in Ring Imaging Cherenkov (RICH) counters \cite{4}.
Ypsilantis and Seguinot \cite{5} proposed a combined aerogel+gas, mirror-focused
RICH counter for the LHC-B experiment at CERN.
The HERMES team  \cite{6} constructed and operated such a
ring imaging detector at DESY. The present paper reports on experimental 
investigation of an aerogel based RICH detector not requiring mirrors
i.e. of the proximity focusing type. Such a detector is being considered
in connection with a possible upgrade of the BELLE particle identification
system at KEK \cite{7,7-1}.

\section{The experimental set-up}

\begin{figure}
\includegraphics[width=9.5cm]{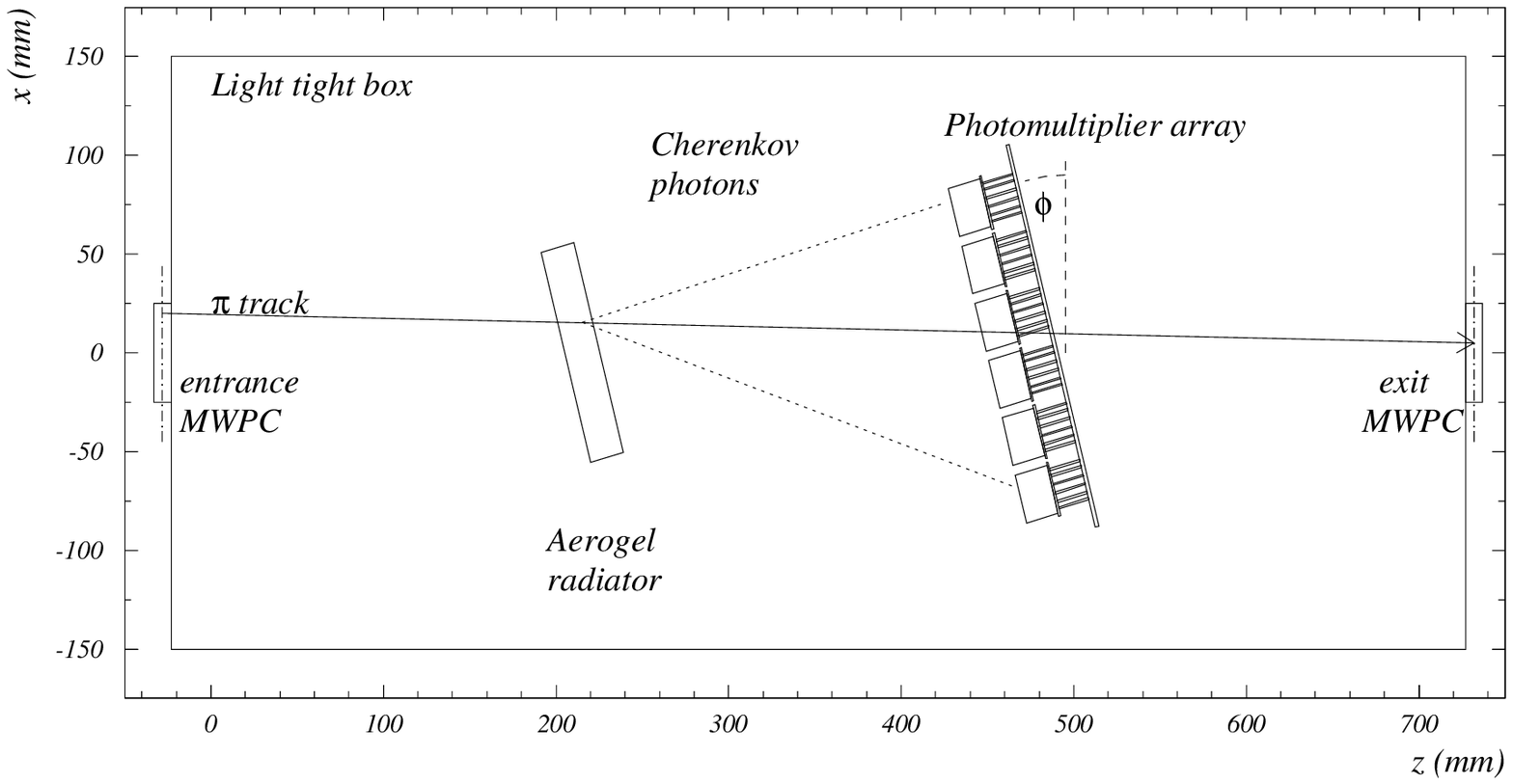}
\caption{The experimental set-up}
\label{fig1}
\end{figure}
Initial tests of the apparatus with cosmic rays were reported recently \cite{8}.
The present paper describes measurements and results obtained with the
$\pi~2$ beam at KEK. A beam particle - pion, muon or
electron - traversing the apparatus is signaled by two $5 \times 5$~cm$^2$
scintillation counters which
determine the time of arrival.
Two CO$_2$ gas Cherenkov counters produce
signals only upon the passage of electrons so these signals could be
used either to select or to exclude electrons.

The aerogel radiator and the position sensitive, single photon detector
are contained in a light tight box (Fig.~\ref{fig1}), of which the entrance 
and exit sides each have a multiwire proportional chamber for measuring
the track of the incident particle. These $5 \times 5$~cm$^2$ MWPC's,
with 15 $\mu$m diameter, gold-plated tungsten anode wires at 2~mm pitch
and with 90\% Ar + 10\% CH$_4$ gas flow, are read out by delay lines on the
x and y cathode wires.

After passing through the entrance MWPC, the charged particle hits 
the aerogel radiator in which it emits Cherenkov photons. Measurements have
been made mainly with 2~cm thick aerogel slabs of $n=1.029$, $n=1.05$ 
and $n=1.07$, produced by the method described in \cite{9}.
The position sensitive detector of Cherenkov photons is situated 17-29~cm
downstream of the aerogel, depending on the refractive index value of the
specific aerogel.                        
The detector is a $6 \times 6$
array of 16 channel multianode photomultiplier tubes 
(Hamamatsu PMTs type R5900-00-M16 with borosilicate window \cite{10})
at 30~mm pitch. The sensitive surface of the M16 PMT is divided into 16 
($= 4 \times 4$) channels, each covering $4.5 \times 4.5$
mm$^2$. It follows that only 36\% of the detector area is occupied
by the photosensitive channels, the rest being dead space.
 The photon detection system and the aerogel radiator tile may be rotated 
around an axis
perpendicular to the beam direction, enabling measurements of angular 
acceptance i.e. measurements of the number of detected Cherenkov photons 
as a function of the charged particle incident angle.

The PMTs are plugged into voltage divider boards inside the light tight
box with signals passing through connectors to the readout system
located outside the box. The PMT anode signals are first 
discriminated and then recorded by CAMAC multihit, multichannel TDCs,
for which the common STOP is provided by the scintillation counter signals. 
The TDC information is stored for later analysis in a personal computer.

As only 196 readout channels were available for the 576 PMT anode outputs,
only part of the system could be read out with the 4.5~mm pixel size.
However, by connecting 4 ($= 2 \times 2$) anodes to one readout channel, 
the entire system could be read out with 9~mm pixel size.

\section{Measurement and results}
\label{MR}

\begin{figure}
\includegraphics[width=9cm]{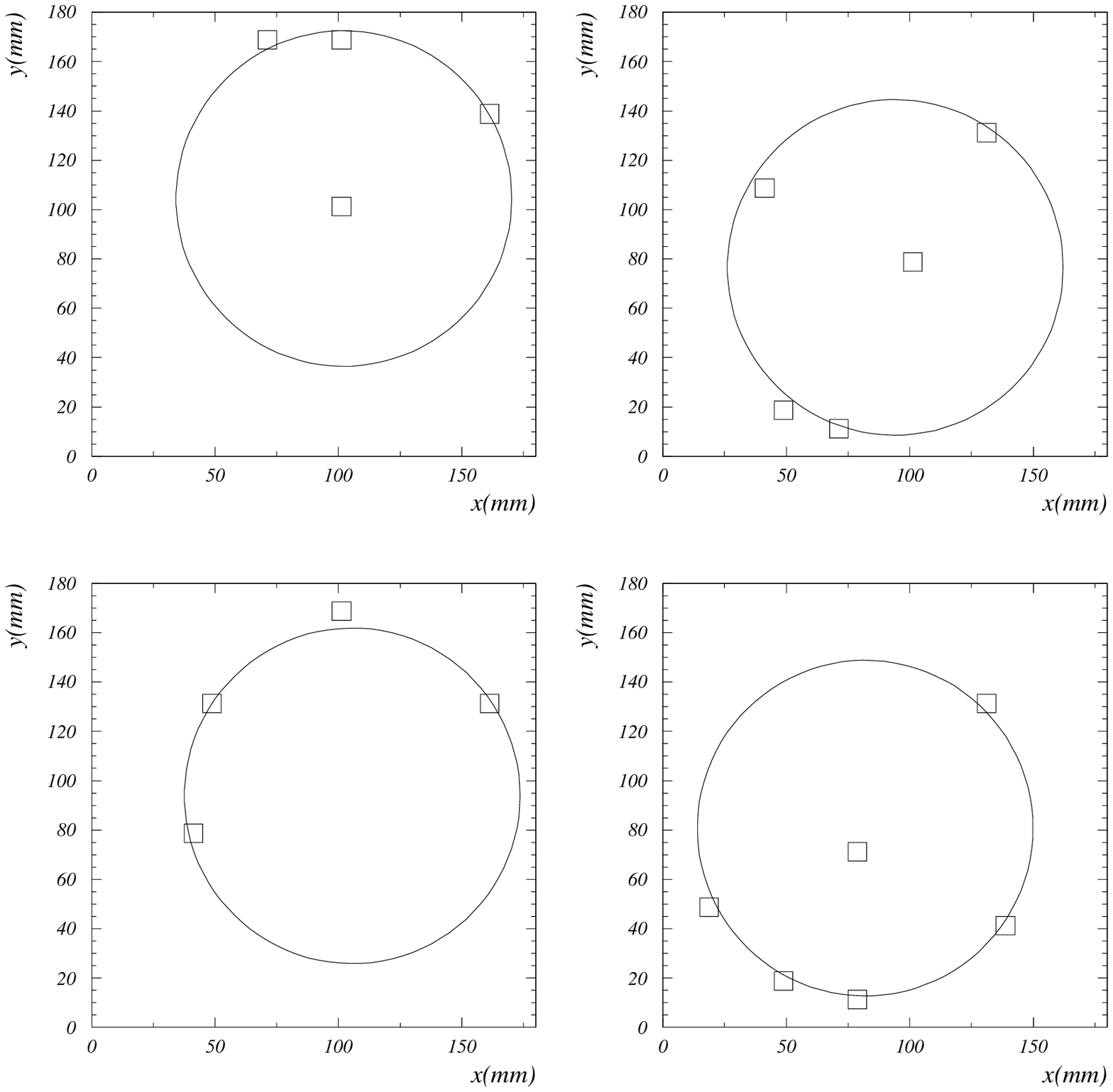}
\caption{Some examples of event hit patterns. The circle corresponds to the
Cherenkov ring of a 3~GeV/$c$ pion given by the measured track position.}
\label{fig2}
\end{figure}
\begin{figure}
\vbox{
\includegraphics[width=8cm]{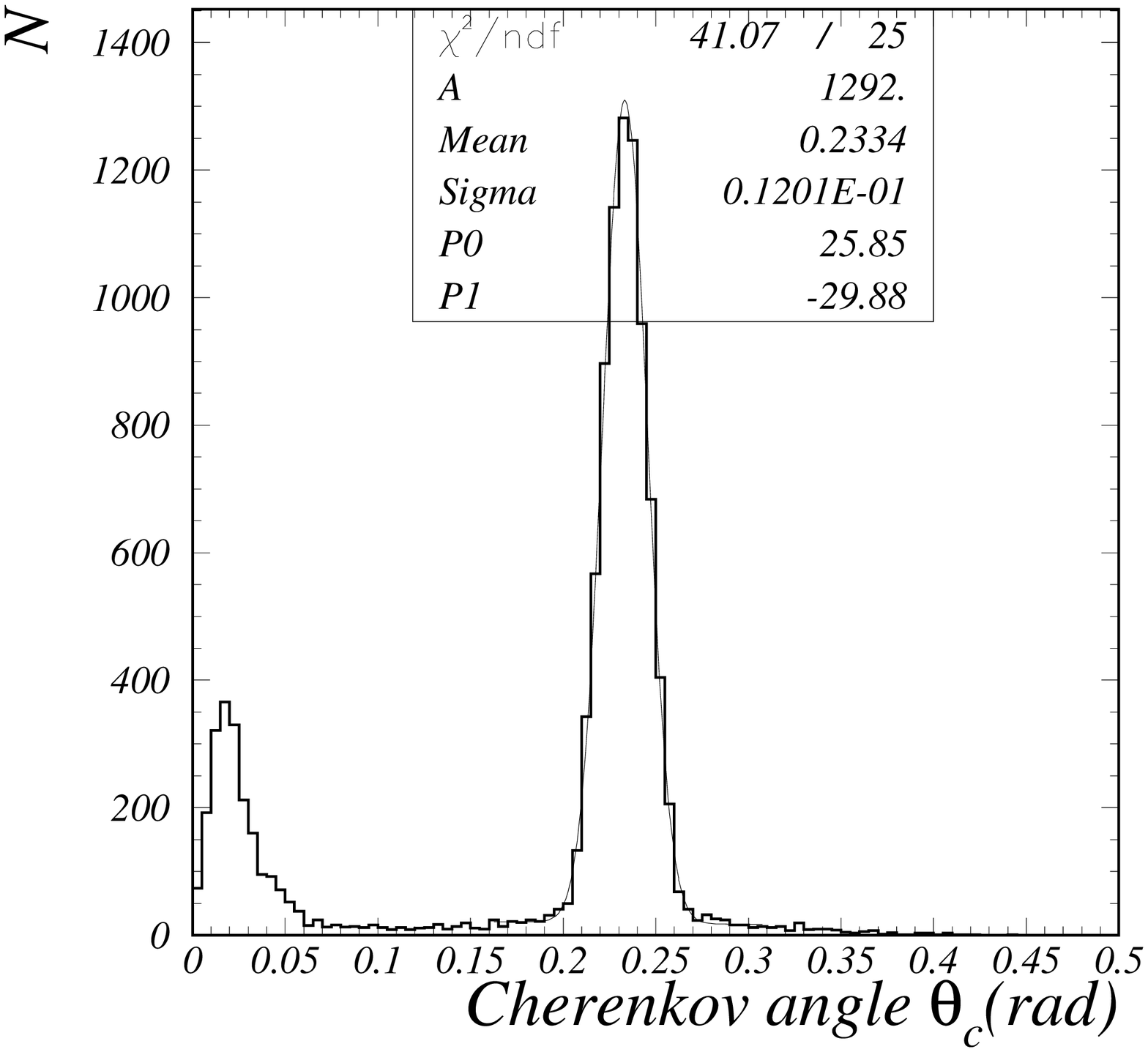}
}
\caption{The distribution of Cherenkov angles measured for  
 3~GeV/$c$ pions,radiating in  an  $n=1.029$ aerogel radiator.
}
\label{fig3-0}
\end{figure}
\begin{figure}
\vbox{
\includegraphics[width=9cm]{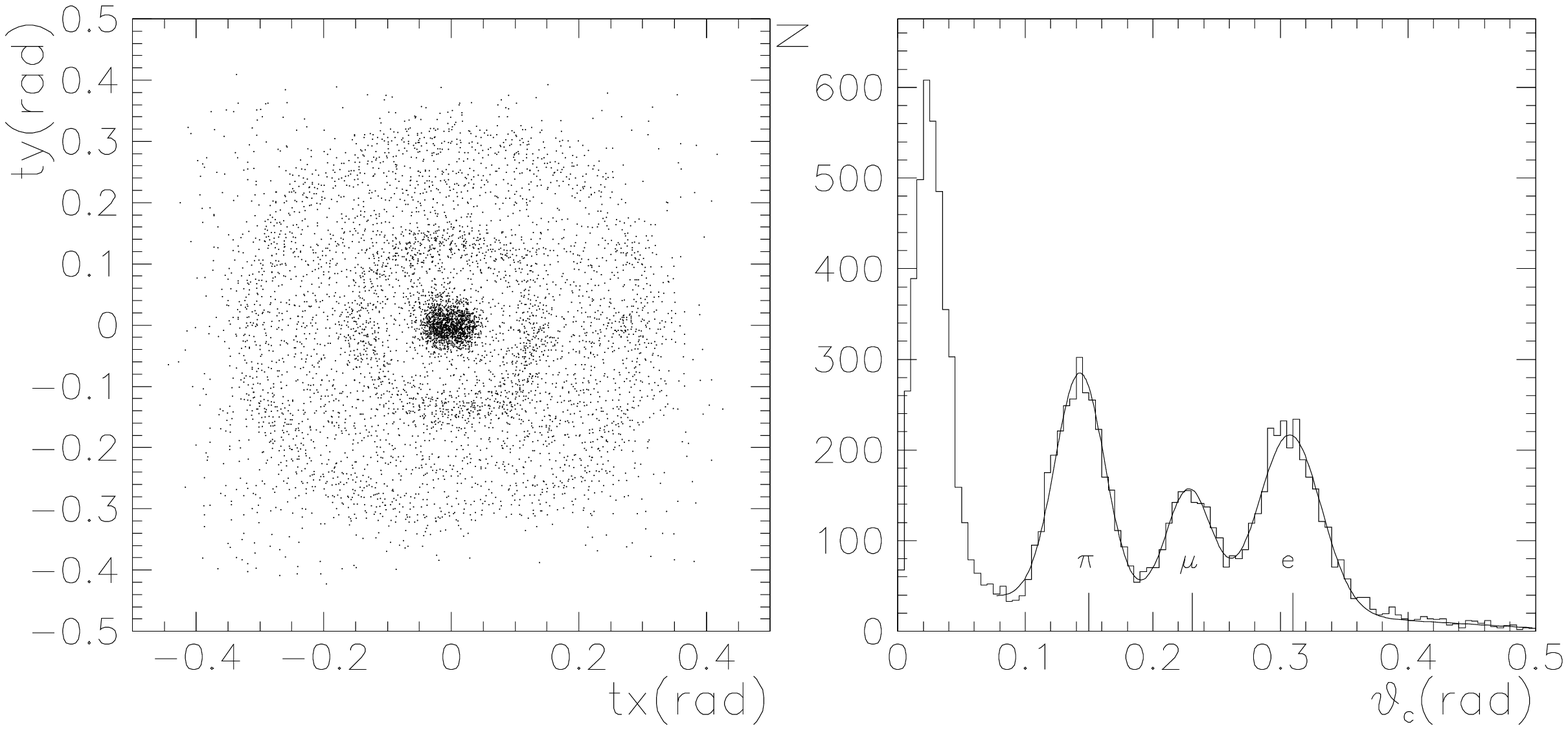}
\includegraphics[width=9cm]{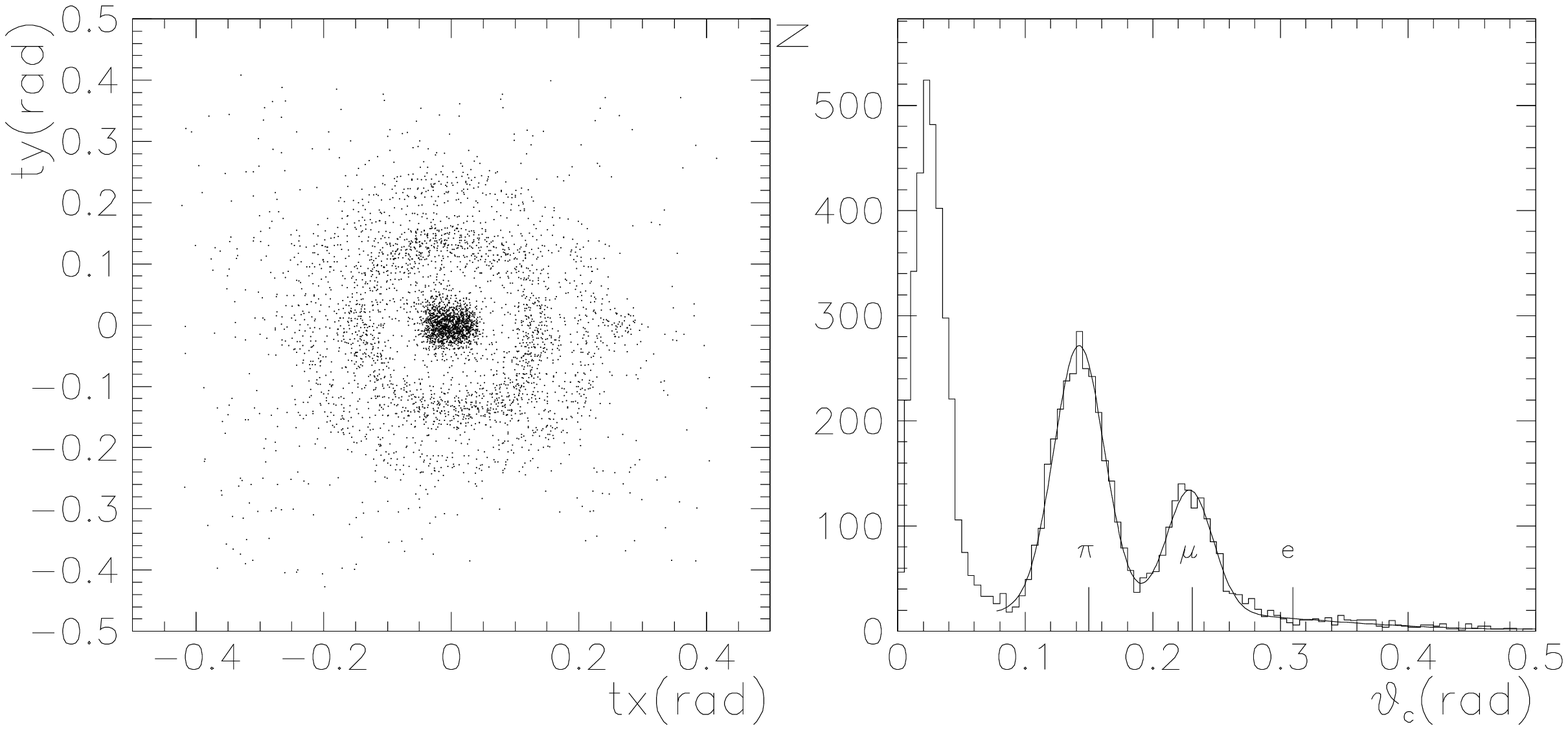}
}
\caption{Distribution of hits in the Cherenkov $x,y$ space (left), 
and versus the Cherenkov polar angle (right)
for 0.5~GeV/$c$ beam particles into an  $n=1.05$ aerogel radiator. The
top graphs show all beam particles, while for the lower
graphs, the gas Cherenkov  has been used to veto the electrons.
}
\label{fig3}
\end{figure}

A few typical events are displayed in Fig.~\ref{fig2}. From the photon hit position
and the measured direction of the incident charged particle,
the  Cherenkov angle is calculated. Accumulated distributions of hits,
depending on their Cherenkov angles, are plotted in Figs.~\ref{fig3-0} and \ref{fig3}. 
Peaks and rings, corresponding to pions, muons and electrons, are clearly visible.
Signals from the gas Cherenkov counters may be used for either selecting or 
excluding electrons. Fitting these distributions
with Gaussian peaks and linear backgrounds yields the average values
and standard deviations of the measured Cherenkov angles.

The main contributions to the resolution in Cherenkov angle 
as determined from a single photon (standard
deviations of the peaks in distributions of Figs.~\ref{fig3-0} and 
\ref{fig3}) come from pixel
size and from uncertainty in the emission point. For normal incidence
of tracks the first contribution could be estimated as 
$\sigma _{pix} = d \cdot $cos$^2 \theta_{ch} /X \sqrt{12}$,
where d is the pixel size, $\theta_{ch}$ is the Cherenkov angle 
and $X$ is the distance from aerogel to detector.
The second contribution is 
$\sigma _{emp} = L \cdot$ sin$ \theta _{ch} \cdot $cos$ \theta_{ch} 
/X \sqrt{12} $, 
where $L$ is the aerogel thickness.
The uncertainty in the track direction  is negligible at 3~GeV/$c$,
but increases at lower momenta. The error due to dispersion
in the radiator (chromatic error) should also be negligible,
but  contributions could arise due to possible non-uniformities
of the aerogel (position variations in refractive index),
non-flat aerogel surface, forward scattering of photons etc. 

\begin{figure}
\includegraphics[width=9cm]{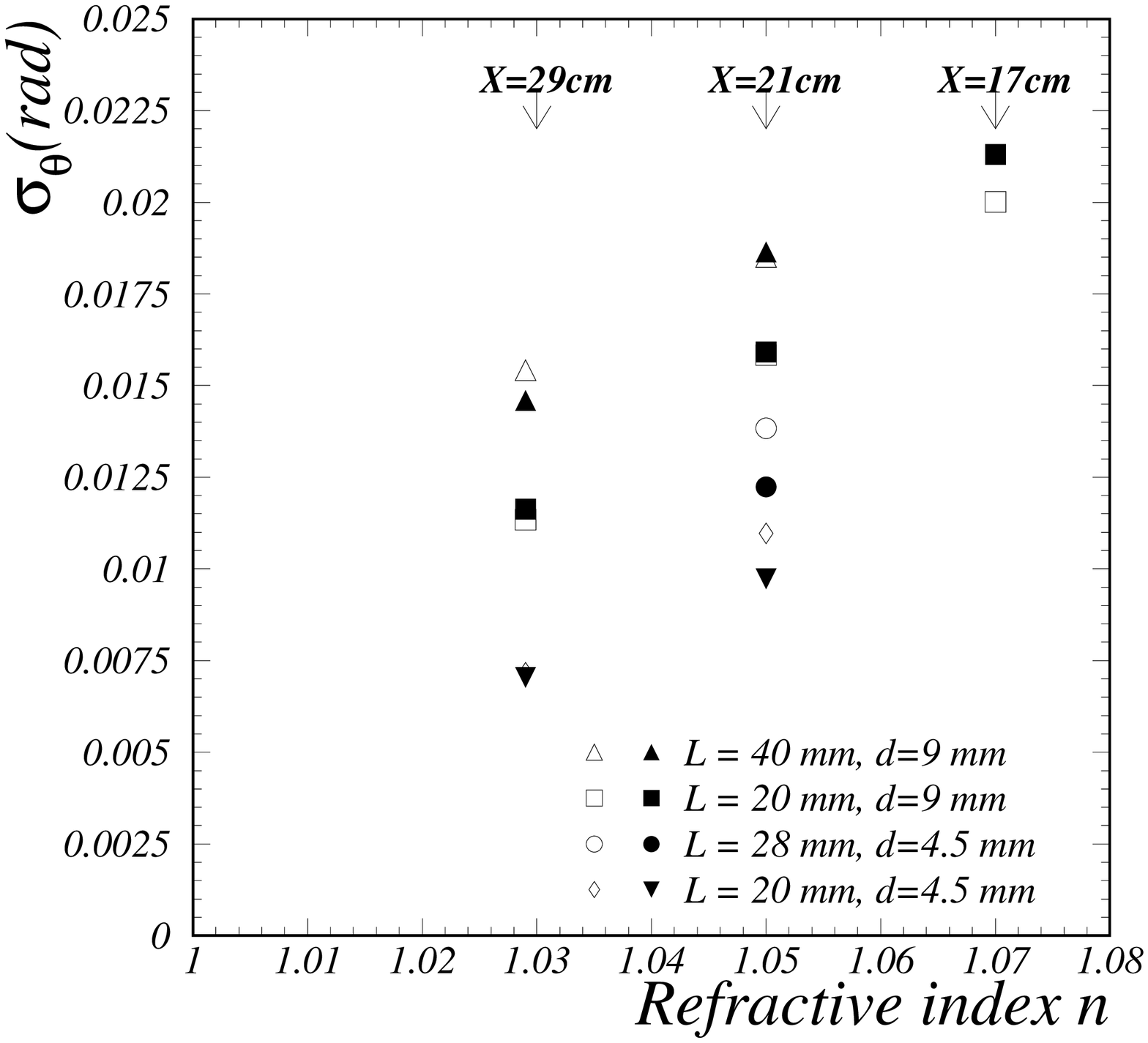}
\caption{ The resolution i.e. the standard deviation of the single Cherenkov
photon angular distribution for different values of the detector parameters,
for a 3~GeV/$c$ pion beam.
Full symbols correspond to the measured values, empty ones to estimates
of contributions from pixel size and emission point uncertainty only.
X is the radiator-to-photon-detector distance, L is the radiator thickness
and d is the photon detector pixel size.}
\label{fig4}
\end{figure}

The measured Cherenkov angle resolution, i.e. the standard deviation
of peaks in distributions of photon hits versus the value of 
aerogel refractive index, is shown in Fig.~\ref{fig4} for 3~GeV/$c$ pions. 
Different data points in the figure refer to different values of 
parameters such as the radiator thickness, the
radiator-to-photon-detector distance and the photon detector pixel size.
The measured values are represented by full symbols, with different
symbol shapes indicating different combinations of parameter values. 
Using the above expressions
for the contributions of pixel size and emission point uncertainty and
summing them in quadrature, one obtains estimates for the resolution,
represented with empty 
symbols in Fig.~\ref{fig4}. It may be seen from the figure that such estimates give
a good approximation to the measured resolution.

\begin{figure}
\includegraphics[width=9cm]{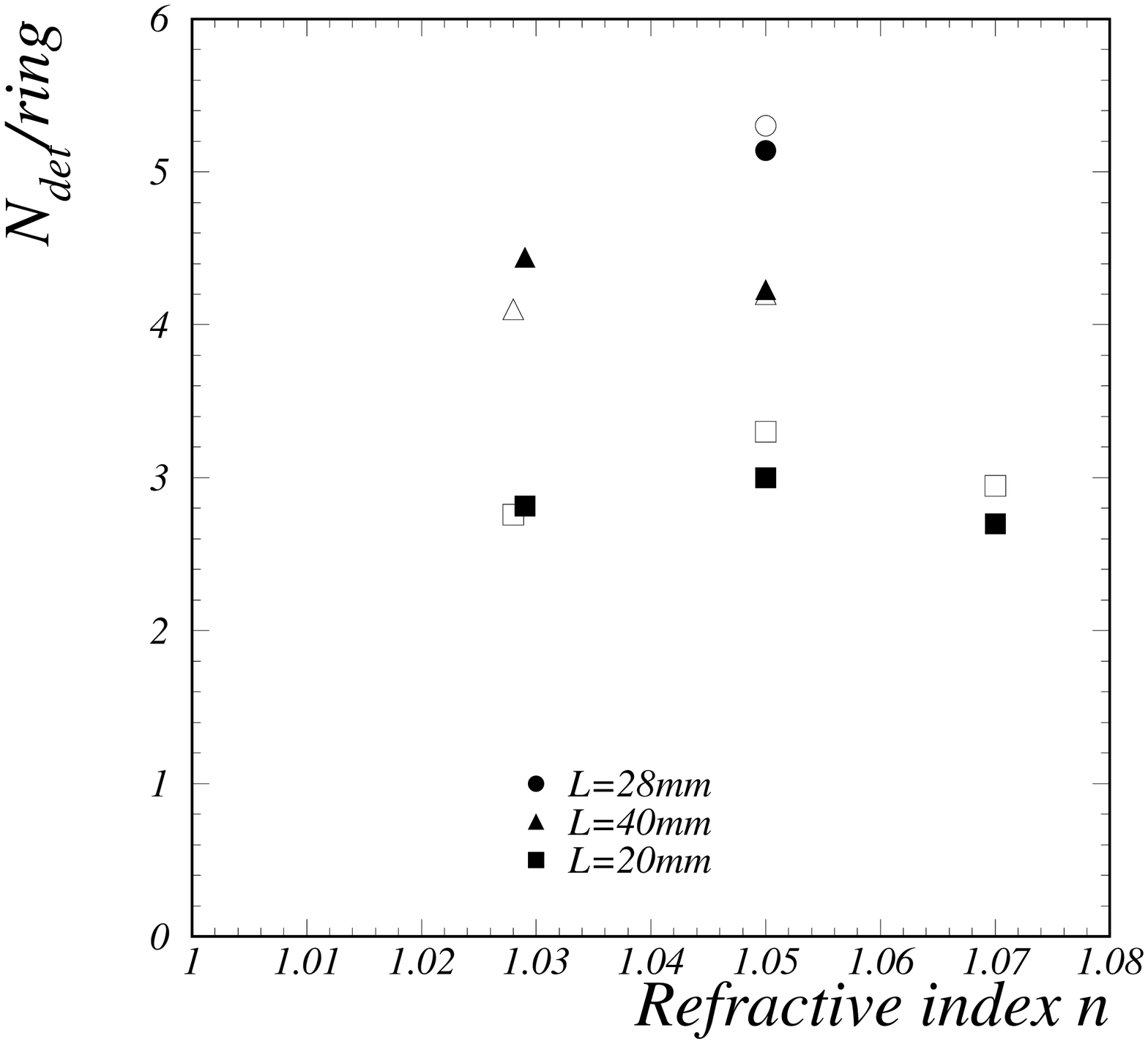}
\caption{Number of detected photons per Cherenkov ring for different aerogels,
for a 3~GeV/$c$ pion beam. L is the aerogel radiator thickness.
Full symbols correspond to the measured values, empty ones to estimates.}
\label{fig5}
\end{figure}
\begin{figure}
\includegraphics[width=9cm]{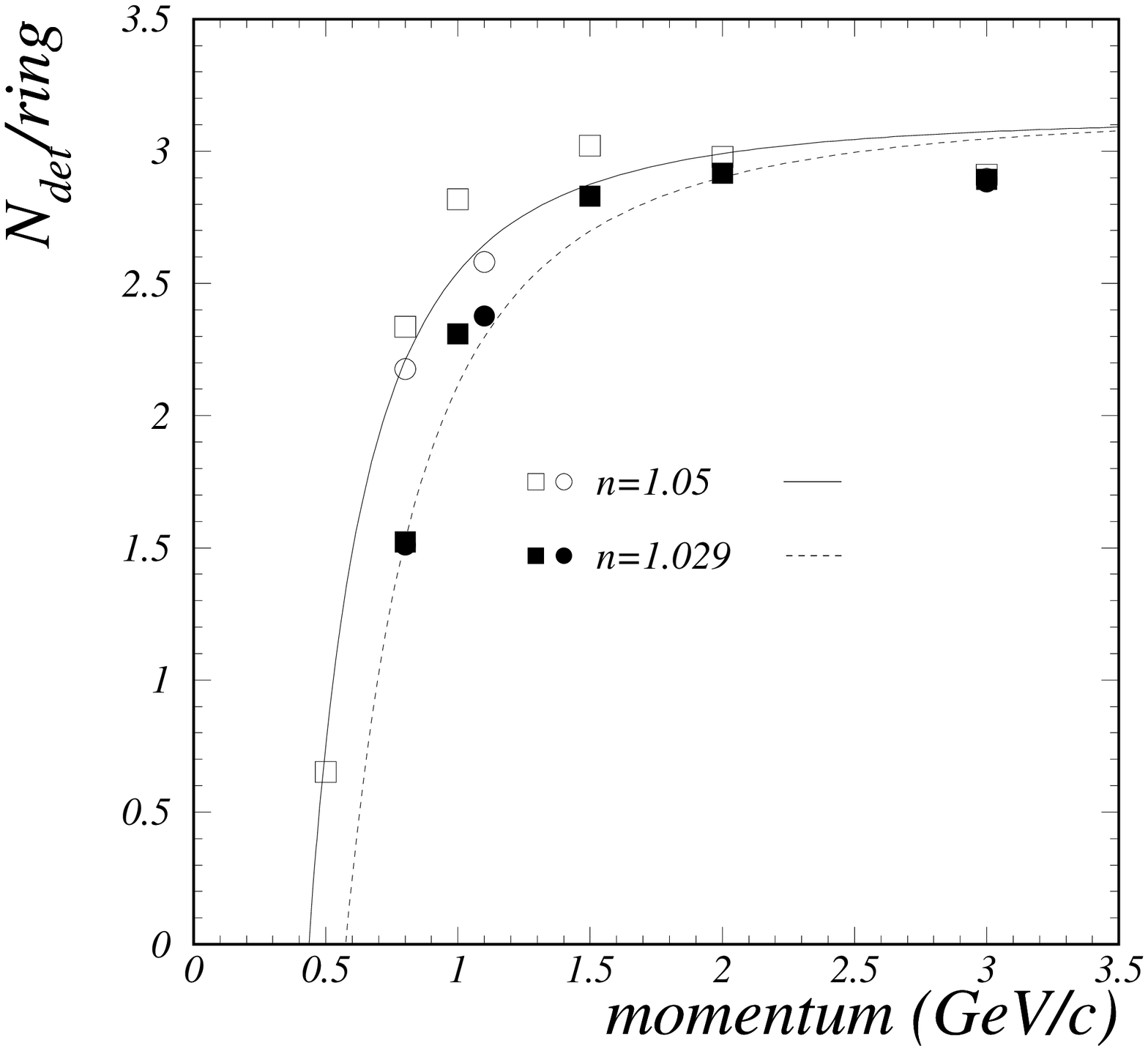}
\caption{Number of detected photons per Cherenkov ring depending on the charged
particle momentum for two different aerogel radiators, with $n=1.05$ 
(open symbols) and $n=1.029$ (full symbols). Squares and circles correspond to 
different granularities of the photon detector. The curves are fits to 
corresponding data.}
\label{fig6}
\end{figure}
The other important parameter of a RICH counter is the number of detected
photons per incident charged particle. This is usually parametrised as
$N_{det} = N_0 \cdot L \cdot sin^2 \theta _{ch}$, where $\theta _{ch}$
is the Cherenkov angle, L is the radiator thickness and $N_0$ is a
figure of merit depending on the radiator and system transparency, 
geometrical acceptance of photons (area and angle), quantum efficiency, 
photoelectron collection efficiency etc. 
Due to Rayleigh scattering, the aerogel
transparency has a strong wavelength dependence in the region of R5900-M16 PMT
quantum efficiency, so one may expect a sensitivity of the number of
detected photons on the particular aerogel sample i.e. on the production
procedure.
The number of detected photons per Cherenkov ring is shown in Fig.~\ref{fig5}
for  3~GeV/$c$ pions.
First one notices that the number of photons does not increase with
refractive index as may be expected for $\beta=1$ particles; 
$N_{det} \propto$ sin$^2 \theta _{ch} = 1 - 1/n^2$. 
Then it is also obvious that the 4~cm
thick aerogel radiator does not produce two times as many photons
as does the 2~cm thick aerogel. And finally we see that the 2.8~cm thick 
aerogel tile \cite{11}, produced in a different process \cite{11-1},
 yields more Cherenkov photons than the 4~cm thick aerogel \cite{9}.
That a higher refractive index of the aerogel sample does not necessarily
produce more photons, is observed also in Fig.~\ref{fig6}.
Although the threshold for $n=1.05$ is reached, as expected, at lower
particle momenta than for $n=1.029$, the saturated number of detected
photons per Cherenkov ring is more or less the same for both radiators.
The above discrepancies can be well understood, 
and have been estimated from the known aerogel attenuation lengths and 
the response of the counter (Fig.~\ref{fig5}). The attenuation lengths
at 400~nm for the samples of Fig.~\ref{fig5} are 
36~mm, 15~mm and 7~mm for the aerogel samples with  $n=1.029$,  
$n=1.05$ and  $n=1.07$, respectively,
and 36~mm for the 28 mm thick Novosibirsk sample with $n=1.05$ \cite{11}.

\begin{figure}
\includegraphics[width=9cm]{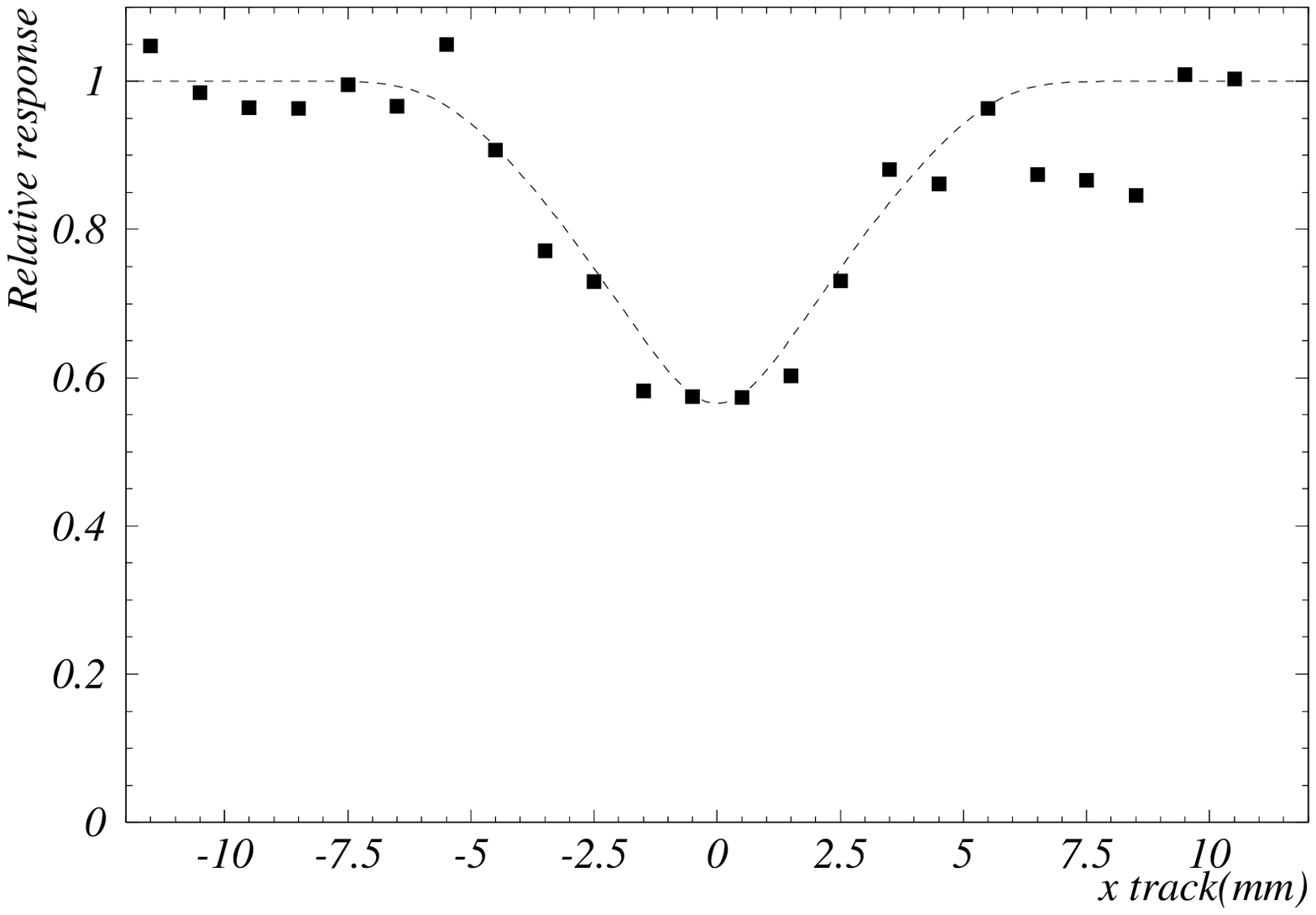}
\caption{The relative 
number of Cherenkov photons detected on the ring, as a function 
of the charged particle impact point distance from the side wall of the
aerogel tile. The values shown are  normalized to the
measured values in case of a single tile covering the full range.
Also shown is the estimate of a simple model (dashed curve).}
\label{fig7}
\end{figure}

It has been already noted by the HERMES group \cite{6}, that a loss of Cherenkov
photons occurs at the side wall boundaries between adjacent aerogel tiles.
We have confirmed this finding by measuring the number of photons on the
Cherenkov ring as a function of the distance of the charged particle impact
point from the boundary between two tiles. The measurement is shown in Fig.~\ref{fig7},
where a dip is seen at the tile boundary $x = 0$~mm. In order to eliminate other
geometrical factors, like for example the acceptance of the photon detector,
the measured yield was normalized to the yield obtained with
one tile covering the entire range. The result clearly
indicates the reduction of yield when the charged particle is closer than
about 5~mm to the boundary of a 2~cm thick $n=1.05$ aerogel tile. It is worth noting that
a simple model, where all photons hitting the boundary between the two tiles get lost,
accounts for most of the observed dependence.

\section{Conclusions}
\label{Co}

We have constructed and tested a proximity focusing RICH detector module with
aerogel as radiator and multianode PMTs as position sensitive detectors 
of individual Cherenkov photons. The measured values of the basic parameters
i.e. the single photon Cherenkov angle resolution and the number of photons
detected per Cherenkov ring, look promising. The resolution is in
relatively good
agreement with estimates based on pixel size and emission point uncertainty.
The number of detected photons, however is sensitive to the particular
aerogel used. It seems that
these differences are due to Rayleigh scattering, which reduces the 
aerogel transparency mainly in the wavelength region of maximal
photocathode     sensitivity.
Photomultipier tubes with a higher fraction of active area, 
possibly combined with 
a light collection system, consisting of lenses or light guides,
are expected to increase the number of detected Cherenkov
photons by reducing the dead space of the photon detector.
The increase in photon yield, however, is in the latter case at the expense of  
an increase in the effective pixel size, so a compromise, optimizing
the final resolution of the charged particle Cherenkov angle should be found.

The information obtained from the results of the present tests suggests that
a proximity focusing aerogel RICH as required by the BELLE particle
identification upgrade is feasible, so investigations of optimal
detector parameters are being continued.

\end{document}